\documentclass[twocolumn,english,aps,reprint,superscriptaddress,showpacs,floatfix]{revtex4}
\usepackage[T1]{fontenc}
\usepackage[latin9]{inputenc}
\usepackage{textcomp}
\usepackage{amssymb}
\usepackage{graphicx}

\makeatletter

\newcommand{\lyxdot}{.}

\@ifundefined{textcolor}{}
{%
 \definecolor{BLACK}{gray}{0}
 \definecolor{WHITE}{gray}{1}
 \definecolor{RED}{rgb}{1,0,0}
 \definecolor{GREEN}{rgb}{0,1,0}
 \definecolor{BLUE}{rgb}{0,0,1}
 \definecolor{CYAN}{cmyk}{1,0,0,0}
 \definecolor{MAGENTA}{cmyk}{0,1,0,0}
 \definecolor{YELLOW}{cmyk}{0,0,1,0}
}

\usepackage{babel}

\usepackage{babel}

\usepackage{babel}

\usepackage{babel}

\makeatother

\usepackage{babel}
\begin{document}

\title{High harmonic imaging of ultrafast many-body dynamics in strongly
correlated systems}

\author{R. E. F. Silva}
\email{silva@mbi-berlin.de}

\affiliation{\emph{Max-Born-Institut, Max Born Strasse 2A, D-12489 Berlin, Germany}}

\author{Igor V. Blinov}

\affiliation{\emph{Russian Quantum Center, Skolkovo 143025, Russia}}

\affiliation{\emph{Moscow Institute of Physics and Technology, 9 Institutsky lane,
Dolgoprudny, Moscow region 141700, Russia}}

\author{Alexey N. Rubtsov}

\affiliation{\emph{Russian Quantum Center, Skolkovo 143025, Russia}}

\affiliation{\emph{Department of Physics, Moscow State University, 119991 Moscow,
Russia}}

\author{O. Smirnova}

\affiliation{\emph{Max-Born-Institut, Max Born Strasse 2A, D-12489 Berlin, Germany}}

\affiliation{\emph{Technische Universitaet Berlin, Ernst-Ruska-Gebaeude, Hardenbergstr.
36A,10623, Berlin, Germany}}

\author{M. Ivanov}
\email{mikhail.ivanov@mbi-berlin.de}

\affiliation{\emph{Max-Born-Institut, Max Born Strasse 2A, D-12489 Berlin, Germany}}

\affiliation{\emph{Blackett Laboratory, Imperial College London, South Kensington
Campus, SW7 2AZ London, United Kingdom}}

\affiliation{\emph{Department of Physics, Humboldt University, Newtonstrasse 15,
12489 Berlin, Germany}}
\begin{abstract}
This Letter brings together two topics that, until now, have been
the focus of intense but non-overlapping research efforts. The first
concerns high harmonic generation in solids, which occurs when intense
light field excites highly non-equilibrium electronic response in
a semiconductor or a dielectric. The second concerns many-body dynamics
in strongly correlated systems such as the Mott insulator. Here we
show that high harmonic generation can be used to time-resolve ultrafast
many-body dynamics associated with optically driven phase transition,
with accuracy far exceeding one cycle of the driving light field.
%inducing the transition. 
Our work paves the way for time-resolving highly non-equilibrium many
body dynamics in strongly correlated systems, with few femtosecond
accuracy. 
\end{abstract}

\pacs{78.47.J\textminus , 71.27.+a, 42.65.Ky}

\maketitle
High harmonic emission provides the frequency domain view of charge
dynamics in quantum systems \cite{krausz2009attosecond}. Complete
characterization of the emitted harmonic light \textendash{} its spectrum,
polarization, and spectral phase \textendash{} allows one to decode
the underlying charge dynamics in atoms and molecules with resolution
$<0.1$ femtosecond (fs), well below a single cycle of the driving
laser field, opening the new field of ultrafast high harmonic imaging
\cite{baker2006probing,lein2007molecular,smirnova2009high,haessler2010attosecond,shafir2012resolving,pedatzur2015attosecond,bruner2016multidimensional,kraus2015measurement}.

Here we bring high harmonic imaging to many-body dynamics in strongly
correlated solids, focusing on an ultrafast phase transition. We consider
the breakdown of the Mott insulating state in the canonical model
of a strongly correlated solid, the Fermi-Hubbard model. We show how
the complex many-body charge dynamics underlying this transition %and the destruction of the insulating state 
are recorded by high harmonic emission, with few fs accuracy.

%Here we bring high harmonic imaging to many-body dynamics in strongly
%correlated solids, focusing on few fsec resolution of an ultrafast
%phase transition. We consider the canonical model of a strongly correlated
%solid, the Fermi-Hubbard model, and the breakdown of the Mott insulating
%state. We show how the correlated charge dynamics associated with
%the transition and the destruction of the insulating state are tracked
%by high harmonic emission with few fsec accuracy.

%High harmonic generation (HHG) has now become an established  
%tool for imaging electron-hole dynamics in atoms and molecules \cite{smirnova2009high, smirnova2009strong, haessler2010attosecond, 
%leeuwenburgh2013high, shafir2012resolving, pedatzur2015attosecond, bruner2016multidimensional, kraus2015measurement}. 
%%Being determined by the Fourier transform of light-induced polarization,  
%spectral range 

Combining high harmonic generation in bulk solids \cite{ghimire2011observation,vampa2014theoretical,Hohenleutner2015,langer2016lightwave,luu2015extreme}
with robust techniques for characterizing the emitted light \cite{Hohenleutner2015,langer2016lightwave,luu2015extreme}
has demonstrated the capabilities of high harmonic imaging in solids.
Recent results include the demonstration of the dynamical Bloch oscillations
\cite{schubert2014sub,luu2015extreme}, the reconstruction of the
band structure in ZnO \cite{vampa2015all}, the visualization of strong
field-induced effective band structure \cite{hawkins2015effect},
and direct visualization of the asymmetric charge flow driven by MIR
fields \cite{you2016anisotropic}. Yet, all of these studies have
been confined to systems well described by single-particle band structure,
and single-particle pictures have dominated the analysis \cite{ghimire2011observation,higuchi2014strong,Hohenleutner2015,vampa2014theoretical,luu2015extreme,hawkins2015effect}.
The role of electron-electron correlation has been relegated to empirically
introduced (and unusually short, only a few fs) relaxation times \cite{vampa2014theoretical,Hohenleutner2015}.

%fields 
%to excite, control, and resolve 
%highly non-equilibrium  electron dynamics
%in the condensed phase and to use high harmonic emission 

%\begin{figure*}
%\begin{centering}
%\includegraphics[width=0.4\textwidth]{FIGURES/hhg} \includegraphics[width=0.45\textwidth]{FIGURES/harmonic_vs_U\lyxdot N_PARTICLES\lyxdot 12} 
%\par\end{centering}
%\caption{\label{fig:fig1}HHG spectrum for several values of $U/t_{0}$ from
%a pulse with $E_{0}=10$ MV/cm, with a frequency of 32.9 THz, 10 optical
%cycles of duration and a $\sin^{2}$ envelope. In (a) it is shown
%$U/t_{0}=0,\,0.1,\,1,\,5$. In (b) it is shown the spectrum for $U/t_{0}$
%from $0$ to $10$. The lower and upper red lines are lower and upper
%bounds of the energetic separation from the groundstate and the single
%doublon-hole excited states and are given by $\Delta$ and $\Delta+8t_{0}$,
%respectively. }
%\end{figure*}

\begin{figure*}
\begin{centering}
\includegraphics[width=0.32\textwidth]{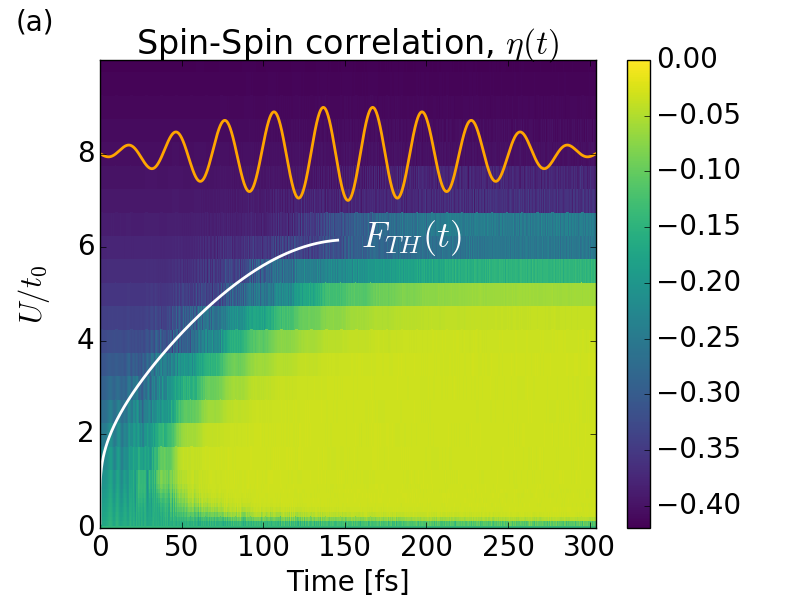}
\includegraphics[width=0.32\textwidth]{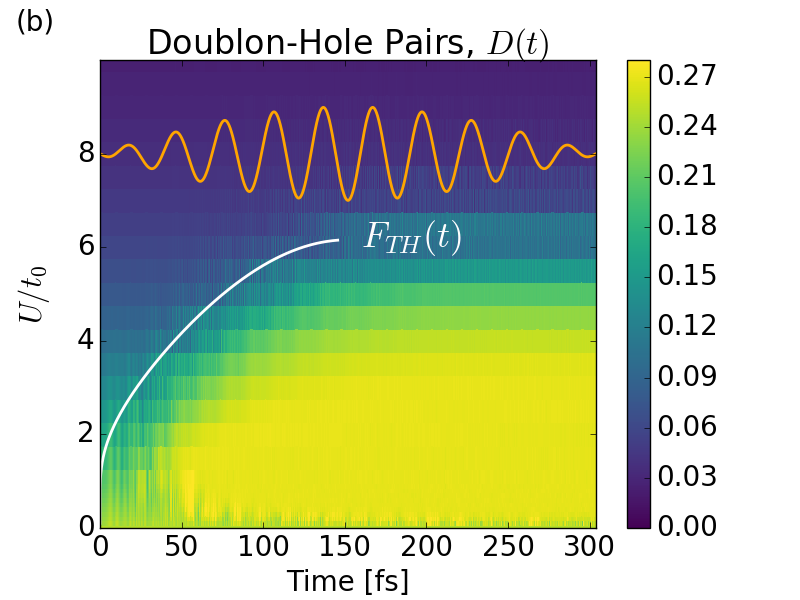}
\includegraphics[width=0.32\textwidth]{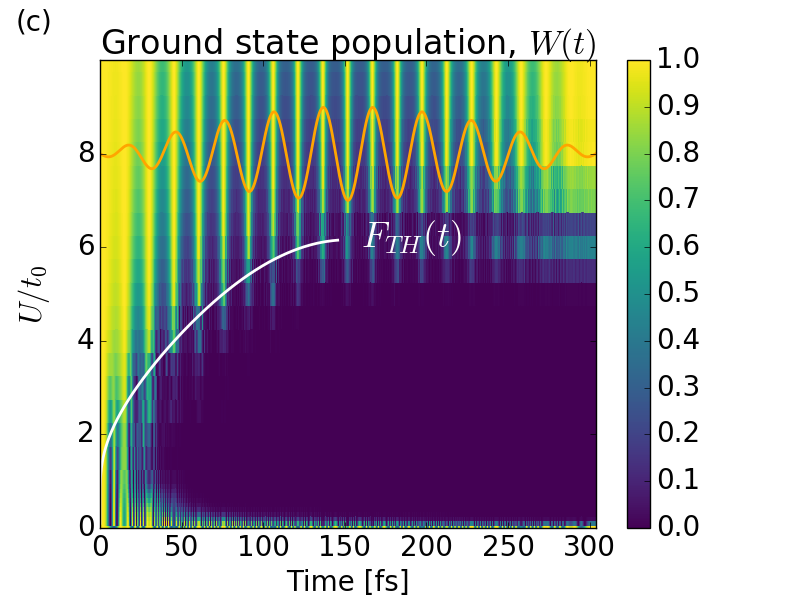} 
\par\end{centering}
\caption{\label{fig:fig1} Time-resolved light-induced breakdown in the Mott
insulator. Time-resolved spin-spin correlation (a) and the number
of doublon-hole pairs (b) for a range of $U/t_{0}$. The pulse has
peak strength $F_{0}=10$ MV/cm, frequency 32.9 THz, and a 10-cycle
long $\sin^{2}$ envelope. The red curve shows the time when the threshold
field strength for the transition is reached. The field amplitude
is insufficient for the breakdown at $U/t_{0}>6$, in agreement with
$F_{{\rm TH}}$ Eq.(\ref{eq:F_TH}). (c) Abrupt destruction of the
Mott state, shown via $W(t)=|\langle\Psi_{0}|\Psi(t)\rangle|^{2}$,
occurs within a cycle for the fields exceeding the threshold. }
\end{figure*}

Yet, electron-electron correlations go well beyond mere dephasing,
generating rich physics of strongly correlated systems, such as pre-thermalization
and the formation of extended Gibbs ensembles \cite{eisert2015quantum},
superfluid to Mott insulator transition \cite{bloch2008many}, to
name but a few. %Cold atoms in optical lattices
%provide an excellent stage for studying the
%interplay of  single-particle and many-body effects \cite{bloch2008many}, from
%such canonical single-particle effects as Bloch oscillations 
%\cite{dahan1996bloch} and Landau-Zener tunneling \cite{niu1996atomic}, 
%to impurity dynamics and the Anderson's
%orthogonality catastrophe \cite{knap2012time}, to such novel collective effects as pre-thermalization and
%the formation of extended Gibbs ensembles \cite{eisert2015quantum}, to
%direct observation of superfluid to a Mott insulator transition \cite{greiner2002quantum}. 
Studies of non-equilibrium many body dynamics have lead to the concepts
of the dynamical \cite{heyl2013dynamical} and light-induced phase
transitions \cite{Nasu2004,oka2008photoinduced,liu2012terahertz,mayer2015tunneling}.
%Light-induced phase
%transitions are among the most interesting testbeds for studying non-equilibrium
%response of strongly correlated systems \cite{Nasu2004}. 
 In particular, the Mott insulator-to-metal transition was recently
achieved experimentally in VO$_{2}$ \cite{liu2012terahertz,mayer2015tunneling}.
Resolving such transitions with few-femtosecond accuracy remains,
however, elusive. Our results show that high harmonic imaging is ideally
suited to address this challenge, offering detailed view of the underlying
dynamics.

We use one-dimensional Fermi-Hubbard model with half-filling, i.e.
with averaged particle density equal to one per site (see Methods
for details). A particle could freely hop to an adjacent site with
a rate $t_{0}$, yielding the metallic state of the system. Hopping
can be obstructed by another particle already residing on the adjacent
site, via the energy $U$ of the repulsive on-site interaction. In
the strong coupling limit $U\gg t_{0}$, the Mott insulating ground
state has short-range antiferromagnetic order \cite{Gebhard1997mottbook}
(the electron spins at the adjacent sites tend to be anti-parallel).
The elementary charge excitations, called doublon-hole pairs \cite{oka2012nonlinear,hubbard_1d_book},
are separated by an optical gap $\Delta$. Deep into repulsive regime
($U\gg t_{0}$), $\Delta$ scales linearly with $U$. We focus on
this regime, and consider the driving field $\omega_{L}\ll\Delta(U)$
in the mid-IR range, with $\omega_{L}=32.9$ THz identical to that
in recent experiments \cite{Hohenleutner2015}, and a modest peak
amplitude $F_{0}=10$ MV/cm. The hopping rate is set to $t_{0}=0.52$
eV to mimic $\mathrm{Sr_{2}CuO_{3}}$ \cite{oka2012nonlinear}, and
$U$ is varied to demonstrate the trends and the general nature of
our conclusions. Details of our simulations are described in the Methods
section.

%The transition between metallic and insulating states has been demonstrated for
%atoms trapped in an optical lattice \cite{greiner2002quantum,bloch2008many}
%by changing the ratio $U/t_{0}$, accomplished by varying the intensity
%of the optical field creating the lattice.
%Smoothly tuning  $U/t_{0}$ is harder in solids. One can, however,
%apply external field to assist tunneling. 
%In the strong coupling limit
%$U/t_{0}\gg 1$ 
%The state is strongly correlated and has antiferromagnetic 
%short order:  the electron spins
%at the adjacent sites are anti-parallel. 

To induce and resolve the Mott transition, we apply a light pulse
where the field amplitude $F_{0}(t)$ increases smoothly. %As it crosses the threshold , it induces an abrupt transition: the breakdown of the insulating ground state is accompanied by the formation of the paramagnetic liquid-like state 
This field can excite the doublon-hole pairs, which play the role
of carrier chargers. The density of doublon-hole pairs may change
during the pulse, depending on the field amplitude $F_{0}$. As $F_{0}$
crosses the threshold $F_{TH}$, the density of charge carriers exceeds
critical value, leading to the breakdown of the Mott insulator \textendash{}
the system becomes conducting. The transition is followed by the destruction
of local magnetic order and a paramagnetic liquid-like state is formed
\cite{oka2008photoinduced,oka2012nonlinear}. Smooth variation of
$F_{0}(t)$ during the pulse allows us to track the transition as
a function of time, with high harmonic response providing sub-cycle
accuracy (see below). The transition is mathematically similar to
strong-field ionization in atoms \cite{oka2012nonlinear}. In particular,
the parameter $\gamma=\hbar\omega_{L}/\xi F_{0}$ (where $\xi$ is
the correlation length \cite{oka2012nonlinear}) serves as the analogue
of the Keldysh adiabaticity parameter \cite{krausz2009attosecond}.
In the `tunnelling' regime $\gamma\ll1$ the threshold field is \cite{oka2012nonlinear}
\begin{equation}
F_{{\rm TH}}=\Delta/2e\xi\label{eq:F_TH}
\end{equation}

%We find that high-frequency emission is synchronized with the onset of phase transition.
As the insulator-to-metal transition is marked by the increased density
of charge carriers and the destruction of short-range magnetic order,
we will characterize the state of the system via the two parameters
describing the charge and spin degrees of freedom: the next-neighbor
spin-spin correlation function 
\begin{equation}
\eta=\frac{1}{L}\left\langle \sum_{j=1}^{L}\vec{S}_{j}.\vec{S}_{j+1}\right\rangle 
\end{equation}
and the average number of doublon-hole pairs per site, 
\begin{equation}
D=\frac{1}{L}\left\langle \sum_{j=1}^{L}c_{j,\uparrow}^{\dagger}c_{j,\uparrow}c_{j,\downarrow}^{\dagger}c_{j,\downarrow}\right\rangle 
\end{equation}
Here $j$ labels the site, up-down arrows the spin, $L=12$ is the
number of cites, $c^{\dagger},c$ are the creation and annihilation
operators.

The destruction of short range antiferromagnetic order during the
transition is shown in Fig. 1 (a): within a cycle, the spin-spin correlation
function drops to nearly zero (limited by the finite size of the system).
The second signature of the transition is the rise in the number of
doublon-hole pairs, Fig.1 (b), which is linked to the loss of spin-spin
correlation (compare Figs. 1 (a,b)). After the transition, the system
reaches a photo-induced saturated state \cite{oka2012nonlinear} and
the number of pairs remains constant. The abrupt nature of the transition
is shown in Fig. 1(c): for fields crossing $F_{{\rm TH}}$ the overlap
probability with the initial state $W(t)=|\langle\Psi_{0}|\Psi(t)\rangle|^{2}$
drops to zero within a laser cycle, stressing the need for sub-cycle
resolution.

Naturally, the rise of optical charge excitations has to manifest
in the optical response. Indeed, we find that the transition is accompanied
by very characteristic high-harmonic emission, see Fig.2.

\begin{figure*}
\begin{centering}
\includegraphics[width=0.4\textwidth]{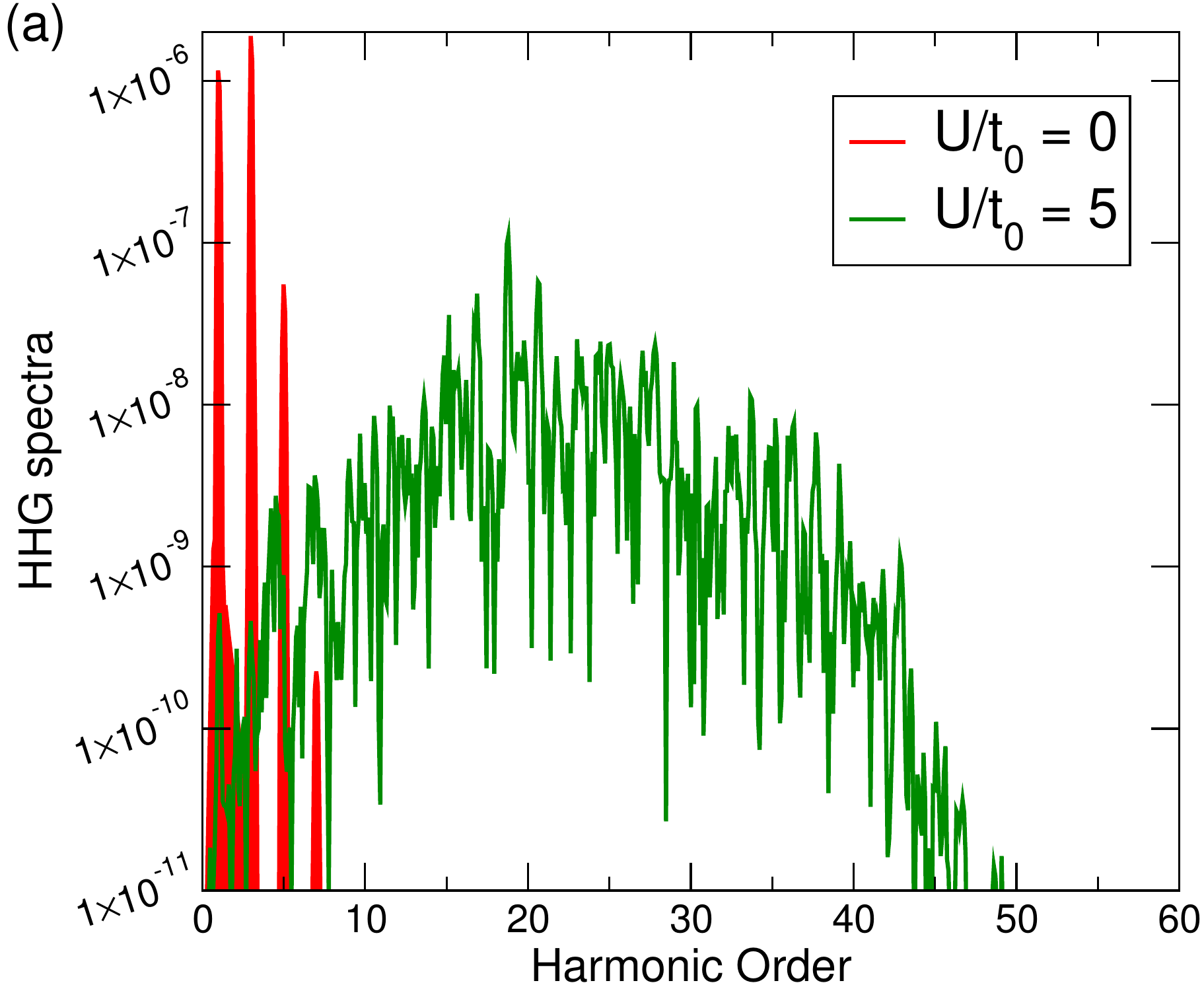} \includegraphics[width=0.45\textwidth]{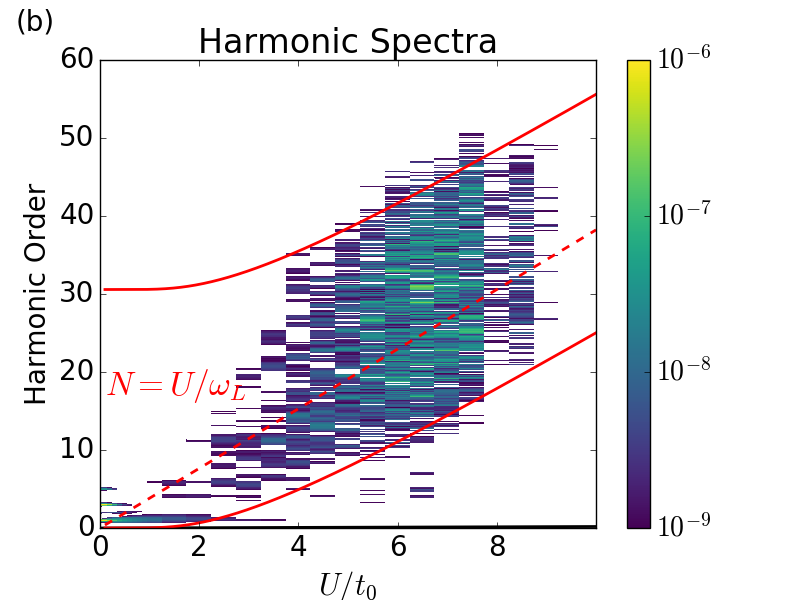}
\includegraphics[width=0.42\textwidth]{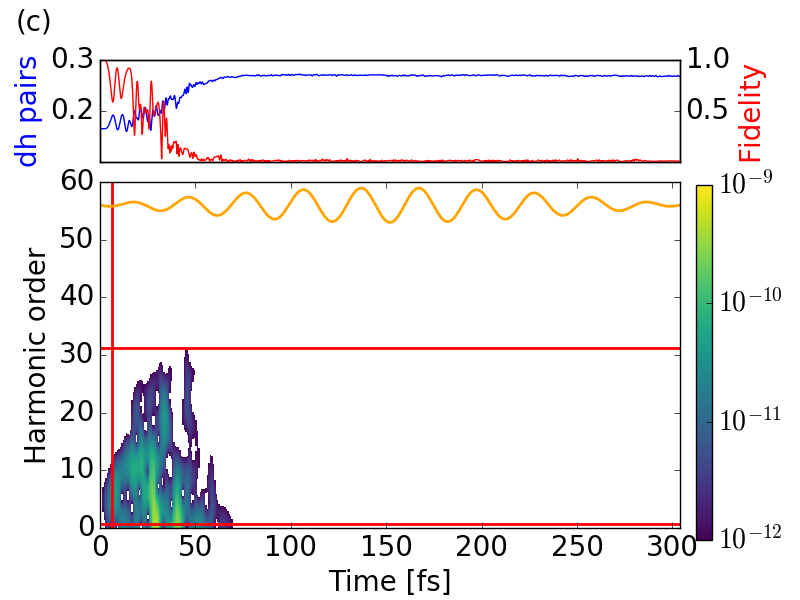} \includegraphics[width=0.42\textwidth]{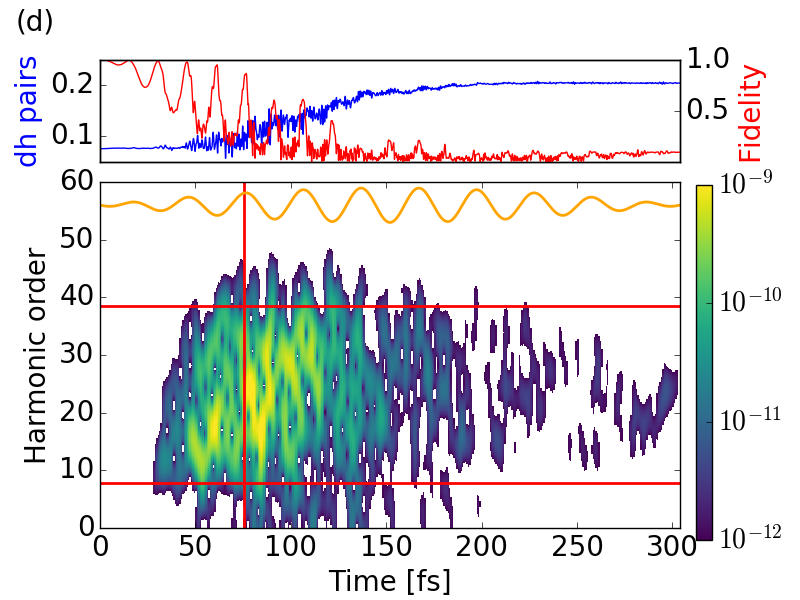} 
\par\end{centering}
\caption{\label{fig:fig2} High harmonic spectroscopy of light-induced transition
in a strongly correlated system. (a) High harmonic spectra for the
conducting state, $U/t_{0}=0$ (red), and the insulating state at
$U/t_{0}=5$ (green). (b) High harmonic spectrum as a function of
$U/t_{0}$. Note dramatic change in the emission spectrum in the strong
coupling limit $U\gg t_{0}$: low harmonics are absent, the emission
peaks at the characteristic energy of doublon-hole excitation $\hbar\Omega\sim U$,
shown with a dashed line. Gabor profiles (c,d) of the harmonic signal
for $U/t_{0}=2$ (c) and $U/t_{0}=5$ (d). The vertical red line shows
when the amplitude $F_{0}(t)$ exceeds the threshold field $F_{{\rm TH}}$
for this value of $U/t_{0}$. The two horizontal red lines are $\Delta$
and $\Delta+8t_{0}$ for this $U/t_{0}$. The top shows the average
number of doublon-hole pairs per site (blue) and the decay of the
insulating state (fidelity $\Xi(t)=|\langle\Psi_{0}|\Psi(t)\rangle|$,
red). }
\end{figure*}

Fig. \ref{fig:fig2} (a) shows harmonic spectra for two different
values of $U/t_{0}$. In the conducting limit $U/t_{0}=0$ the emission
is typical for single band tight binding model \cite{ghimire2011observation},
demonstrating clear low-order Bloch oscillation-type harmonics associated
with intra-band current (the intra-band harmonics). As expected, the
harmonics are narrow and well defined. In the case of $U/t_{0}\gg1$,
the spectrum is quite unusual.

First, the intra-band harmonics are strongly suppressed, in stark
contrast with systems described by single-particle band structures
(e.g. \cite{ghimire2011observation,schubert2014sub,vampa2014theoretical,Hohenleutner2015,langer2016lightwave}).
Second, for the same $F_{0}$, the spectrum becomes much broader and
shifts towards orders $N\sim U/\omega_{L}$. This characteristic change
is summarized in Fig. \ref{fig:fig2} (b), where we scan $U/t_{0}$
to demonstrate the trend. For $U/t_{0}\gg1$, the spectrum peaks near
the characteristic energies of doublon-hole excitation. In the half-filled
system, the first allowed charge excitations are states with single
doublon-hole pairs with energies between $\Delta$ and $\Delta+8t_{0}$
\cite{hubbard_1d_book,oka2012nonlinear}. Fig.2(b) shows that these
excitations are the ones responsible for the harmonic emission. Indeed,
their range, shown with red lines in Fig. \ref{fig:fig2} (b), defines
the lower and upper frequencies for the harmonic emission. Thus, the
emission corresponds to the one-photon transition that brings the
excited system back to its initial ground state via the doublon-hole
recombination.

Third, the regular structure of the harmonic lines is lost in the
strong coupling limit. This stands in stark contrast to weakly correlated
systems, where electron-electron correlation is expected to introduce
fast dephasing, the latter yielding regular, narrow harmonic lines
\cite{vampa2014theoretical}. Figs.2(c,d) clarify the physics responsible
for the irregular structure of the spectrum.

%Higher-order excitations above $\Delta+8t_{0}$
%are not coupled to the ground state by single-particle transitions, and thus
%the corresponding emission is suppressed.
%Fig. \ref{fig:fig2} (b) shows that, for sufficiently high $U/t_{0}$,
%in the tunnelling regime, the new many-body state of
%the system created by the photo-induced transition 
%does not support Bloch-like oscillations of doublon-hole pairs,
%which would have generated strong low-order harmonics familiar from
%one-electron-type excitations in a conventional conduction band \cite{oka2012nonlinear}.
%Loosely speaking, this happens because all quasi-momenta states for the doublon-hole pairs
%are occupied upon the tunneling transition.

%Fig. \ref{fig:fig2} shows the harmonic spectra for a
%wide range of $U/t_{0}$ values. We observe that the harmonic signal,
%for values of $U/t_{0}\ge2$, is located between $\Delta$ and $\Delta+8t_{0}$.

%in which the HHG cutoff scales linearly with the strength field, and
%suggests that in strongly correlated systems the mechanism of harmonic
%generation strongly relies on the electron correlation. In our results,
%the cutoff is located at $\Delta+8t_{0}$ and the harmonic signal
%is only relevant at energies greater than the Mott gap.

Figs. \ref{fig:fig2} (c,d) show the time profile of the harmonic
emission, obtained via the Gabor transform (see Methods). Note that
complete time-domain reconstruction of the emitted harmonic light
with $\sim1-2$ fs accuracy is fully within the available experimental
technology \cite{Hohenleutner2015}, for the same laser parameters
as in our calculation. %Time-resolving the emission captures these processes
%with resolution determined by the overall spectral bandwidth, e.g.
%$\sim1$ fs for $U/t_{0}=5, t_{0}=0.52$ eV. 
%Figs. \ref{fig:fig2} (c,d) show that the harmonic emission is
%synchronized with the transition and occurs mostly while the number
%of doublon-hole pairs is changing. 
We see that (i) the onset of the harmonic emission is synchronized
with the breakdown of the insulating Mott state and the rise in the
number of doublon-hole pairs, and (ii) the fall of the emission follows
the depletion of the insulating state (Fig.2(c,d)), following the
fidelity $\Xi(t)=|\langle\Psi_{0}|\Psi(t)\rangle|$. The Gabor profiles
in Fig. 2(c,d) show that the emission takes about 50-70 fsec, i.e.
only about 1-2 cycles of the driving field. The temporal restriction
of the emission to a couple of cycles of the driving field explains
the lack of clear peaks at odd harmonics. The complexity of the spectrum
affirms the strongly aperiodic many-body dynamics, in contrast to
the periodic intraband motion in the limit $U/t_{0}\ll1$ (see Fig.2(a)).
The top panels in Figs. 2(c,d) confirm the conclusions drawn from
Fig. 2(b): the emission relies on the coherence created between the
Mott insulator ground state and the doublon-hole states. This is why
it starts when the doublon-hole pairs are created and ends when the
ground, Mott insulator state, is destroyed.

The lack of low-order harmonics after the phase transition leads to
another important conclusion: the new many-body state created by the
photo-induced transition does not support Bloch-like oscillations
of doublon-hole pairs. Indeed, these would have generated strong low-order
harmonics familiar from one-electron-type excitations in a conventional
conduction band. Loosely speaking, this happens because all quasi-momenta
states for the doublon-hole pairs are occupied upon the transition
in the tunneling regime \cite{oka2012nonlinear}.

In contrast to high harmonic emission in systems with single-particle
band structure \cite{ghimire2011observation,schubert2014sub,luu2015extreme,vampa2014theoretical},
the cutoff of the harmonic signal associated with a phase transition
in a strongly correlated system does not scale linearly with the electric
field. In our case the harmonic emission has threshold behavior and,
at $F_{0}\geq F_{{\rm TH}}$, covers all energies between $\Delta$
and $\Delta+8t_{0}$ irrespective of the field. We also find that
strong electron-electron correlations do not necessarily lead to the
emergence of regular harmonic spectra with well defined lines, as
expected for weakly correlated systems. Highly irregular harmonic
spectra imply highly aperiodic dynamics, in line with the dramatic
change in the state of the system during a phase transition.

High harmonic generation has been pioneered three decades ago \cite{ferray1988multiple},
evolving from an unusual table-top source of bright, coherent XUV
light to the technological backbone of attosecond science \cite{krausz2009attosecond}
and a unique tool for imaging ultrafast dynamics with attosecond to
few-femtosecond temporal resolution \cite{lein2007molecular}. Yet,
throughout these decades, the analysis of high harmonic generation
has been rooted in effectively single-electron pictures. Our work
is the first to bring fundamental strongly correlated many-body dynamics
squarely into its view.

\begin{acknowledgments}
We gratefully acknowledge fruitful discussions with Dr. Takashi Oka,
Dr. Bruno Amorim, and Dr. Peter Hawkins. M. I. and R. E. F. S. acknowledge
the support from the MURI programme. 
\end{acknowledgments}

\section*{Methods\label{sec:theoretical}}

We study high harmonic generation in the one-dimensional, half-filled
Fermi-Hubbard model by solving the time dependent Schr{ö}dinger's
equation (TDSE) numerically exactly, fully including the electron-electron
correlations in the system interacting with intense light field. We
use the 1D Fermi-Hubbard Hamiltonian \cite{hubbard_1d_book} 
\begin{eqnarray}
\hat{H}\left(t\right) & = & -t_{0}\sum_{\sigma,j=1}^{L}\left(e^{-i\Phi\left(t\right)}c_{j,\sigma}^{\dagger}c_{j+1,\sigma}+e^{i\Phi\left(t\right)}c_{j+1,\sigma}^{\dagger}c_{j,\sigma}\right)\nonumber \\
 & + & U\sum_{j=1}^{L}c_{j,\uparrow}^{\dagger}c_{j,\uparrow}c_{j,\downarrow}^{\dagger}c_{j,\downarrow}.\label{eq:HubbardHamLaser}
\end{eqnarray}
where the laser electric field $F(t)=-dA(t)/dt$ enters through the
time-dependent Peierls phase $eaF\left(t\right)=-d\Phi\left(t\right)/dt$,
$a$ is the lattice constant and $A(t)$ is the field vector potential.
The hopping parameter $t_{0}$ is set to $t_{0}=0.52$ eV to mimic
$\mathrm{Sr_{2}CuO_{3}}$ \cite{oka2012nonlinear}, and $U>0$ is
the on-site Coulomb repulsion. In the calculations, we use the periodic
boundary conditions $c_{j,\sigma}=c_{j+L,\sigma}$ with $L=N=12$,
$N$ being the number of particles, and focus on the $S_{z}=0$ subspace.
Starting at $t=0$ from the ground state of the Hamiltonian, we apply
the pulse with $A\left(t\right)=A_{0}f\left(t\right)\sin\left(\omega_{L}t\right)$
at the carrier wavelength of $9.11\mu$m ($\omega_{L}=32.9$ THz)
and the peak amplitude $F_{0}=\omega_{L}A_{0}=10$ MV/cm. All the
parameters of the pulse are well within the experimental reach. The
pulse has total duration of 10 optical cycles and a $\sin^{2}$ envelope,
and is shown in Fig.1

To compute the harmonic emission, we first use the electric current
operator, defined as \cite{hubbard_1d_book} 
\begin{eqnarray}
J\left(t\right) & = & -ieat_{0}\sum_{\sigma}\sum_{j=1}^{L}\left(e^{-i\Phi\left(t\right)}c_{j,\sigma}^{\dagger}c_{j+1,\sigma}-h.c.\right).\label{eq:CurrentOperator}
\end{eqnarray}
to compute the time-dependent current. The harmonic spectrum is calculated
as the square of the Fourier transform of the dipole acceleration,
$a\left(t\right)=\frac{d}{dt}J\left(t\right)$. Time-resolved emission
is obtained by performing the Gabor (window Fourier) transform with
the sliding window $\exp[-(t-\tau)^{2}/\sigma^{2}]$, $\sigma=\left(3\omega_{L}\right)^{-1}$.

\appendix
%dummy comment inserted by tex2lyx to ensure that this paragraph is not empty%dummy comment inserted by tex2lyx to ensure that this paragraph is not empty%dummy comment inserted by tex2lyx to ensure that this paragraph is not empty%dummy comment inserted by tex2lyx to ensure that this paragraph is not empty
 \bibliographystyle{naturemag}
%%%\bibliography{biblio_hubbard}

\end{document}